\documentclass[preprintnumbers, prd, onecolumn, floatfix, superscriptaddress, nofootinbib]{revtex4-2}

\usepackage{amsmath, amssymb, geometry}
\usepackage{graphicx}
\usepackage{setspace}
\usepackage{subfig}
\usepackage{epsfig}
\usepackage{bm}
\usepackage{float}
\usepackage{dcolumn}
\usepackage{cancel}
\usepackage[colorlinks]{hyperref}
\usepackage[usenames,dvipsnames]{color}
\usepackage{enumitem}
\usepackage{xcolor}

\hypersetup{
	breaklinks=true,
	pdfstartview={FitH},
	colorlinks=true,
	linkcolor=blue,
	citecolor=red,
	filecolor=magenta,
	urlcolor=blue,
	anchorcolor=green,
	linktocpage=true
}

\def\doi{http://doi.org}

\def\and{$and$}

\newcommand{\be}{\begin{equation}}
	\newcommand{\ee}{\end{equation}}
\newcommand{\ban}{\begin{eqnarray*}}
	\newcommand{\ean}{\end{eqnarray*}}
\newcommand{\ba}{\begin{eqnarray}}
	\newcommand{\ea}{\end{eqnarray}}

\newcommand{\bc}{\begin{center}}
	\newcommand{\ec}{\end{center}}

\begin{document}
	
	\title{Growth Rate Analysis in \boldmath$f(R,L_m)$ Gravity: A Comparative Study with \boldmath$\Lambda$CDM Cosmology}
	
	\author{G. K. Goswami}
	\email{gk.goswami9@gmail.com}
	\affiliation{Department of Mathematics, Netaji Subhas University of Technology, New Delhi-110 078, India}
	\author{J. P. Saini}
	\email{jps@mmmut.ac.in}
	\affiliation{Vice Chancellor, Madan Mohan Malviya University of Technology, Gorakhpur, Uttar Pradesh, India}
	\begin{abstract}
		We investigate the evolution of cosmic structures within the framework of modified gravity, specifically focusing on theories described by the function \( f(R, L_m) \), where \( R \) is the Ricci scalar and \( L_m \) is the matter Lagrangian. This class of models introduces a non-minimal coupling between geometry and matter, leading to modifications in the dynamics of density perturbations. 
		
		We derive the linear growth equation and compute the observable growth rate \( f\sigma_8(z) \), which is directly accessible from redshift-space distortion (RSD) data. Using recent observational constraints from galaxy surveys such as eBOSS and DESI, we perform a comparative analysis between predictions from \( f(R, L_m) \) gravity and the standard \( \Lambda \)CDM model. 
		
		Our results indicate that while \( \Lambda \)CDM remains broadly consistent with current data, the \( f(R, L_m) \) framework can accommodate subtle deviations in structure growth, offering a possible resolution to existing tensions in large-scale structure observations. We also outline the implications of our findings for future surveys, including Euclid and LSST.
	\end{abstract}
	\maketitle
	
	\section{Introduction}
	
	Over the past two decades, a wide range of cosmological observations—including Type Ia supernovae (SNe Ia) \cite{Riess1998, Perlmutter1999}, cosmic microwave background (CMB) anisotropies \cite{Planck2018}, and baryon acoustic oscillations (BAO) \cite{Eisenstein2005}—have provided compelling evidence that the Universe is undergoing a phase of late-time accelerated expansion. The standard model of cosmology, the $\Lambda$CDM model, explains this phenomenon by invoking a cosmological constant $\Lambda$ and cold dark matter (CDM) in the context of a spatially flat Friedmann–Lemaître–Robertson–Walker (FLRW) universe.
	
	Despite its remarkable success in accounting for a broad array of observational data, the $\Lambda$CDM model is not without challenges. It faces several conceptual issues, such as the fine-tuning and coincidence problems associated with the cosmological constant. In addition, persistent observational tensions—such as the discrepancy in the measured values of the Hubble constant $H_0$ from early and late Universe probes \cite{Riess2021, DiValentino2021}, and the $\sigma_8$ tension in large-scale structure clustering amplitude \cite{Heymans2021}—have prompted the investigation of alternative models.
	
	Among the most promising avenues are \textit{modified gravity theories}, which extend Einstein’s General Relativity by altering the gravitational action. These include $f(R)$ gravity, scalar–tensor theories, $f(T)$ gravity, and others. Such frameworks aim to account for cosmic acceleration without introducing a separate dark energy component, attributing the phenomenon instead to modifications in the geometry of spacetime.
	
	In this work, we investigate a lesser-studied yet theoretically appealing extension known as $f(R, L_m)$ gravity \cite{Harko2010}, in which the gravitational Lagrangian is a function of both the Ricci scalar $R$ and the matter Lagrangian $L_m$. This approach introduces a non-minimal coupling between curvature and matter, leading to modified field equations and novel features in cosmic dynamics. One of its distinctive predictions is the alteration of the growth of matter perturbations, which is crucial for understanding the formation of large-scale structure.
	
	Our primary goal is to explore the evolution of cosmic structures in $f(R, L_m)$ gravity and confront its predictions with observational data. Specifically, we derive the linear growth equation for density perturbations and compute the observable growth rate parameter $f\sigma_8(z)$, which can be directly constrained using redshift-space distortion (RSD) measurements from surveys such as eBOSS and DESI. We also compare the results with predictions from the $\Lambda$CDM model to assess the viability of the modified framework.
	
	\vspace{0.2cm}
	\noindent
	\textbf{Organization of the paper:}
	\begin{itemize}
		\item Section~2 outlines the theoretical formulation of $f(R, L_m)$ gravity and presents the modified Friedmann equations.
		\item Section~3 describes the solutions for the Hubble function and effective energy density.
		\item Section~4 discusses the observational datasets and statistical techniques used for parameter estimation, including MCMC sampling.
		\item Section~5 focuses on redshift-space distortions, defines the $f\sigma_8(z)$ observable, and details its computation in $f(R, L_m)$ gravity.
		\item Section~6 provides a comparison between theoretical predictions and the $f\sigma_8(z)$ observational dataset.
		\item Section~7 concludes with a summary of the results and their implications.
	\end{itemize}
	\section{Theoretical Framework}
	
	We consider a class of modified gravity theories described by the functional form
	\[
	f(R, L_m) = \alpha R + L_m^\beta + \gamma,
	\]
	where \( R \) is the Ricci scalar, \( L_m \) is the matter Lagrangian density, and \( \alpha \), \( \beta \), and \( \gamma \) are constants. This form introduces a non-minimal coupling between geometry and matter, modifying both the background expansion and the evolution of cosmic perturbations.
	
	\subsection{Modified Friedmann Equations and Their Solutions}
	
	The action for \( f(R, L_m) \) gravity is given by~\cite{Harko2010}:
	\[
	S = \int \left[ \frac{1}{2\kappa^2} f(R, L_m) + \mathcal{L}_m \right] \sqrt{-g} \, d^4x,
	\]
	where \( \kappa^2 = 8\pi G \), and \( \mathcal{L}_m \) is the matter Lagrangian density.
	
	Varying the action with respect to the metric \( g_{\mu\nu} \) yields the field equations:
	\begin{align}
		f_R R_{\mu\nu} - \frac{1}{2} f g_{\mu\nu} + (g_{\mu\nu} \Box - \nabla_\mu \nabla_\nu) f_R = 
		\frac{1}{2} f_{L_m} T_{\mu\nu} + \left(1 - f_{L_m}\right) \nabla_\mu \nabla_\nu L_m,
	\end{align}
	where subscripts denote partial derivatives, e.g., \( f_R = \partial f / \partial R \), and \( f_{L_m} = \partial f / \partial L_m \).
	
	In general, the energy-momentum tensor is not conserved, satisfying the relation:
	\[
	\nabla^\mu T_{\mu\nu} = \frac{f_{L_m}}{f_{L_m}} \left( g_{\mu\nu} L_m - T_{\mu\nu} \right) \nabla^\mu \ln f_{L_m}.
	\]
	
	We adopt the functional form \( f(R, L_m) = \alpha R + L_m^\beta + \gamma \), as discussed in~\cite{Myrzakulov2024}, and assume \( L_m = \rho \), where \( \rho \) is the energy density of matter. For a spatially flat FLRW universe, the metric takes the form
	\[
	ds^2 = -dt^2 + a^2(t)(dx^2 + dy^2 + dz^2),
	\]
	leading to the modified Friedmann equations:
	\begin{align}
		3 H^2 &= \frac{1}{2\alpha} \left[(2\beta - 1)\rho^\beta - \gamma\right], \label{eq:fried1} \\
		2\dot{H} + 3H^2 &= -\frac{1}{2\alpha} \left[(1 - \beta)\rho^\beta + \beta \rho^{\beta-1} p + \gamma \right]. \label{eq:fried2}
	\end{align}
	
	Here, \( H = \dot{a}/a \) is the Hubble parameter, and the energy-momentum tensor is modeled as a perfect fluid:
	\[
	T_{\mu\nu} = (\rho + p)u_\mu u_\nu + p g_{\mu\nu}, \quad u^\mu = (0, 0, 0, 1),
	\]
	with a barotropic equation of state \( p = (1 - n)\rho \), where \( n \) is a model parameter.
	
	To express the field equations in terms of redshift \( z \), we use the relations:
	\[
	1 + z = \frac{a_0}{a(t)}, \quad \dot{H} = - (1 + z) H(z) \frac{dH}{dz}.
	\]
	
	Substituting into Eq.~\eqref{eq:fried2} gives the redshift-dependent differential equation:
	\[
	\beta n \left( \gamma + 6\alpha H(z)^2 \right) - 4\alpha (2\beta - 1)(1 + z) H(z) \frac{dH}{dz} = 0.
	\]
	
	Solving this equation yields the Hubble parameter as a function of redshift:
	\[
	H(z)^2 = \left( \frac{\gamma}{6\alpha} + H_0^2 \right)(1 + z)^{\frac{3\beta n}{2\beta - 1}} - \frac{\gamma}{6\alpha}.
	\]
	
	This can be recast in the familiar form:
	\[
	H(z) = H_0 \sqrt{(1 - \lambda) + \lambda (1 + z)^{3(1 + w)}},
	\]
	where the effective parameters are defined as:
	\[
	\lambda = \frac{\gamma}{6\alpha H_0^2} + 1, \quad w = \frac{\beta(n - 2) + 1}{2\beta - 1}.
	\]
	
	Thus, the background cosmological dynamics in this framework are governed by the parameter set \( \{\alpha, \beta, \gamma\} \), or equivalently, \( \{H_0, \lambda, w\} \).
	
	\subsection*{Energy Density as a Function of Redshift}
	
	For a pressureless matter-dominated Universe (i.e., \( n = 0 \)), Eq.~\eqref{eq:fried1} provides the following redshift-dependent energy density:
	\begin{equation} \label{eq:frlm_rho}
		\rho(z) = \left( \frac{\gamma + 6\alpha H^2(z)}{2\beta - 1} \right)^{1/\beta}.
	\end{equation}
	
	This expression is particularly useful when solving the linear perturbation equations for structure formation in the $f(R, L_m)$ framework.
	\section{Observational Constraints}
	
	We constrain the parameters of the $f(R, L_m)$ model—namely, \( H_0 \), \( \lambda \), and \( w \)—using a joint analysis of recent cosmological datasets. These include Hubble parameter measurements, Type Ia supernovae from the Pantheon$^+$ sample, baryon acoustic oscillations (BAO), and the CMB shift parameter.
	
	\subsection{Methodology}
	
	The total chi-square is defined as:
	\begin{equation}
		\chi^2_{\text{total}} = \chi^2_{\text{Hubble}} + \chi^2_{\text{SNe}} + \chi^2_{\text{BAO}} + \chi^2_{\text{CMB}},
	\end{equation}
	where each term represents the contribution from a different observational probe. The theoretical Hubble parameter is given by:
	\begin{equation}
		H(z) = H_0 \sqrt{(1 - \lambda) + \lambda (1 + z)^{3(1+w)}},
	\end{equation}
	and the luminosity distance is computed as:
	\begin{equation}
		d_L(z) = (1 + z) \int_0^z \frac{dz'}{H(z')}.
	\end{equation}
	
	\subsection{Datasets Used}
	
	The analysis utilizes the following datasets:
	
	\begin{itemize}[leftmargin=1.5em]
		\item \textbf{Hubble Parameter Data:} 35 measurements from cosmic chronometers~\cite{Moresco:2022ynk}, providing direct estimates of $H(z)$ across redshift.
		
		\item \textbf{Type Ia Supernovae:} The Pantheon$^+$ sample~\cite{Brout2022}, comprising 1701 SNe Ia, with the full statistical and systematic covariance matrix.
		
		\item \textbf{Baryon Acoustic Oscillations:} The DESI DR2 data~\cite{Wang:2025}, reporting \( D_H(z)/r_d \) and \( D_M(z)/r_d \) measurements in the redshift range \( 0.5 \leq z \leq 2.3 \).
		
		\item \textbf{CMB Shift Parameter:} The Planck 2018 value of the CMB shift parameter, \( R_{\text{obs}} = 1.7492 \pm 0.0049 \), evaluated at \( z_* = 1089.92 \)~\cite{Planck2018}.
	\end{itemize}
	
	\subsection{Statistical Analysis}
	
	We perform parameter estimation via chi-square minimization, followed by a Bayesian Markov Chain Monte Carlo (MCMC) analysis to explore the posterior distributions. In parallel, a neural network-based fit was employed to cross-validate the results. The outputs are presented through corner plots and comparative error bar visualizations.
	
	\subsection{Results and Discussion}
	
	For the $f(R, L_m)$ model, the best-fit parameters at 68\% confidence level are:
	\[
	H_0 = 73.75 \pm 0.16~\mathrm{km\,s^{-1}\,Mpc^{-1}}, \quad
	\lambda = 0.262 \pm 0.007, \quad
	w = -0.005 \pm 0.001.
	\]
	
	For comparison, the corresponding $\Lambda$CDM model yields:
	\[
	H_0 = 73.49 \pm 0.14~\mathrm{km\,s^{-1}\,Mpc^{-1}}, \quad
	\Omega_m = 0.278 \pm 0.006.
	\]
	
	The results indicate that both models support a relatively high Hubble constant \( H_0 \), consistent with local measurements. The $f(R, L_m)$ framework allows for small but potentially significant deviations from $\Lambda$CDM, as encoded in the effective equation-of-state parameter \( w \) and the coupling parameter \( \lambda \).
	
	\begin{figure}[H]
		\centering
		\includegraphics[width=0.45\textwidth]{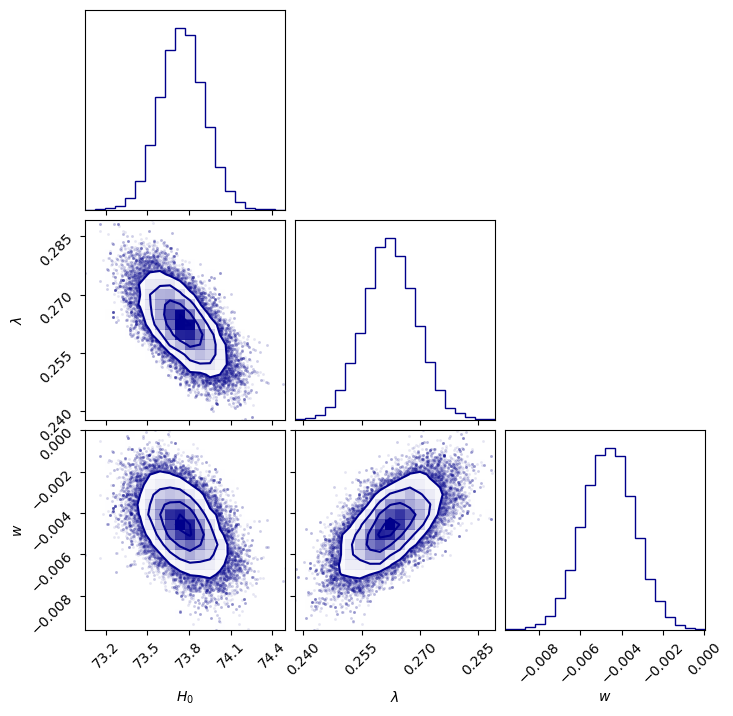}
		\includegraphics[width=0.45\textwidth]{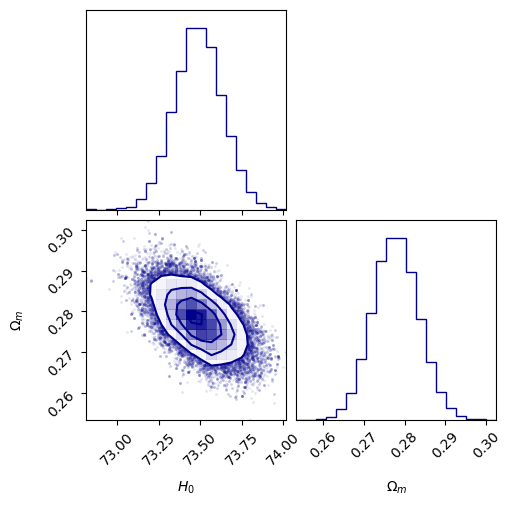}
		\caption{Corner plots showing marginalized posterior distributions and parameter correlations. 
			Left: $f(R, L_m)$ model with best-fit values 
			$H_0 = 73.75^{+0.16}_{-0.16}~\mathrm{km\,s^{-1}\,Mpc^{-1}}$, 
			$\lambda = 0.262^{+0.007}_{-0.007}$, and 
			$w = -0.005^{+0.001}_{-0.001}$.
			Right: $\Lambda$CDM model with 
			$H_0 = 73.49^{+0.15}_{-0.14}~\mathrm{km\,s^{-1}\,Mpc^{-1}}$, 
			and $\Omega_m = 0.278^{+0.006}_{-0.006}$.}
		\label{fig:corner}
	\end{figure}
	
	\begin{figure}[H]
		\centering
		\includegraphics[width=0.75\textwidth]{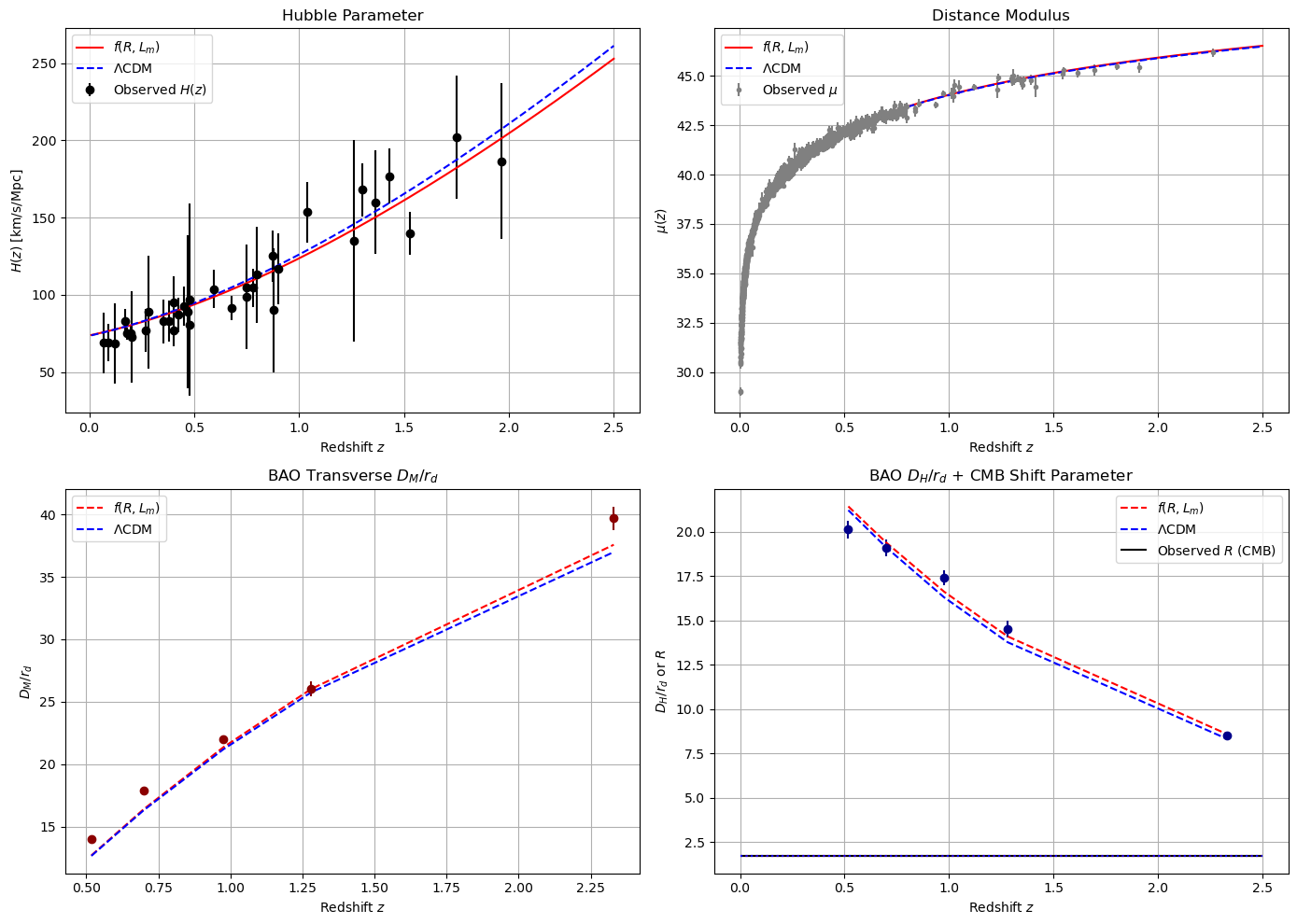}
		\caption{Panel plots comparing theoretical predictions and observational data for Hubble parameter $H(z)$, distance modulus $\mu(z)$, BAO, and CMB shift parameter.}
		\label{fig:fourpanel}
	\end{figure}
	
	\begin{figure}[H]
		\centering
		\includegraphics[width=0.75\textwidth]{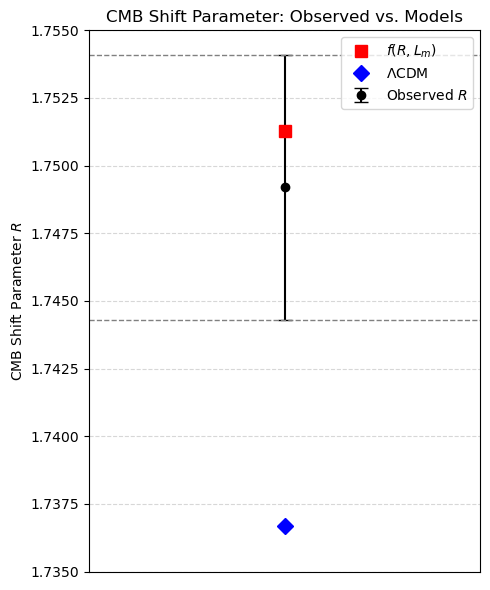}
		\caption{Error bar plots for $H(z)$, $\mu(z)$, BAO, and CMB data. The $f(R, L_m)$ model fits closely with current observations across all redshift bins.}
		\label{fig:errorbars}
	\end{figure}
	\section{Growth of Matter Perturbations in Redshift Space}
	
	The evolution of matter density perturbations, denoted by \( \delta(z) \), is a key diagnostic of structure formation in cosmology. In the standard $\Lambda$CDM framework, the linear growth of sub-horizon perturbations is governed by the second-order differential equation:
	\begin{equation}
		\frac{d^2 \delta}{dz^2} + \left( \frac{d \ln H}{dz} - \frac{2}{1+z} \right) \frac{d\delta}{dz}
		- \frac{3}{2} \frac{\Omega_m (1+z)}{E^2(z)} \delta = 0,
	\end{equation}
	where \( E(z) = H(z)/H_0 \) is the normalized Hubble parameter. This equation encapsulates the interplay between gravitational collapse and cosmic expansion.
	
	To solve this equation numerically, we recast it into a system of two first-order differential equations and integrate using Python’s \texttt{solve\_ivp} routine. Initial conditions are specified at high redshift based on the analytical solution in the matter-dominated era:
	\[
	\delta(z) = \frac{1}{1+z}, \qquad \frac{d\delta}{dz} = -\frac{1}{(1+z)^2}.
	\]
	
	\subsection{Modified Growth in \boldmath\( f(R, L_m) \) Gravity}
	
	For the modified gravity theory defined by
	\[
	f(R, L_m) = \alpha R + L_m^\beta + \gamma,
	\]
	with \( L_m = \rho \), the evolution of perturbations is altered due to the non-minimal coupling between geometry and matter. Under the Newtonian gauge and quasi-static approximation, the perturbed $(0,0)$ component of the field equations yields a modified Poisson equation:
	\[
	\nabla^2 \Phi = \frac{1}{2} G_{\text{eff}} a^2 \rho \delta, \qquad 
	G_{\text{eff}} = \frac{(2\beta - 1)\beta \rho^{\beta - 1}}{\alpha}.
	\]
	
	Substituting this into the Euler and continuity equations leads to the modified growth equation:
	\begin{equation}
		\boxed{
			\ddot{\delta} + 2H \dot{\delta} - \frac{1}{2} G_{\text{eff}} \rho \delta = 0.
		}
	\end{equation}
	
	Transforming this to redshift space via \( \frac{d}{dt} = -H(1+z)\, \frac{d}{dz} \), the numerical solution requires the following functions:
	\begin{align}
		H(z) &= H_0 \sqrt{(1 - \lambda) + \lambda(1 + z)^{3(1 + w)}}, \\
		\rho(z) &= \left( \frac{\gamma + 6\alpha H^2(z)}{2\beta - 1} \right)^{1/\beta}.
	\end{align}
	
	\subsection{Results and Comparison}
	
\begin{figure}[h!]
		\centering
		\includegraphics[width=0.80\textwidth]{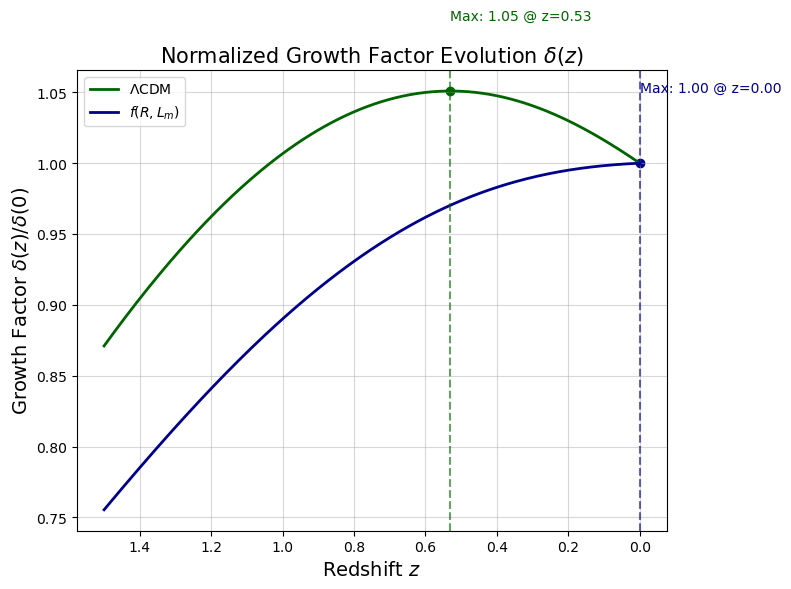}
		\caption{Evolution of the normalized growth factor \( \delta(z)/\delta(0) \) for the $\Lambda$CDM and \( f(R, L_m) \) models. The $\Lambda$CDM model exhibits extended growth up to \( z \approx 0.53 \), reaching a peak of \( \delta/\delta_0 \approx 1.05 \), whereas the \( f(R, L_m) \) model saturates earlier with a peak value of unity at \( z = 0 \).}
		\label{fig:growth_comparison}
	\end{figure}

	Figure~\ref{fig:growth_comparison} illustrates the redshift evolution of the normalized growth factor, \( \delta(z)/\delta(0) \), for both the $\Lambda$CDM and \( f(R, L_m) \) models. In the $\Lambda$CDM case, perturbations continue to grow until \( z \approx 0.53 \), after which they begin to saturate. In contrast, the \( f(R, L_m) \) model exhibits earlier saturation, resulting in a lower growth amplitude at late times.
	
	\begin{table}[h!]
		\centering
		\caption{Comparison of Growth Factor Peaks in $\Lambda$CDM and $f(R, L_m)$ Models}
		\vspace{0.5em}
		\begin{tabular}{lcc}
			\hline
			\textbf{Model} & \textbf{Peak \( \delta(z)/\delta(0) \)} & \textbf{Redshift of Peak} \\
			\hline
			$\Lambda$CDM & 1.05 & 0.53 \\
			$f(R, L_m)$  & 1.00 & 0.00 \\
			\hline
		\end{tabular}
		\label{tab:growth_peak}
	\end{table}
	
	\paragraph{Peak Growth Behavior.} The peak behavior of the growth factor provides a potential observational discriminator between gravity models. While the standard $\Lambda$CDM model allows structure growth to persist until relatively late times, the \( f(R, L_m) \) model leads to earlier stabilization of the matter contrast. This implies a suppression in late-time structure formation and may help explain certain tensions in large-scale structure observations.
	
	\section*{Redshift-Space Distortions and \boldmath$f\sigma_8(z)$}
	
	\textbf{Redshift-Space Distortions (RSD):} In galaxy redshift surveys, the observed redshift combines contributions from the Hubble expansion and the peculiar velocities of galaxies. These peculiar motions introduce anisotropies in the redshift-space distribution—an effect known as \textit{redshift-space distortions} (RSD).
	
	\vspace{0.3cm}
	\textbf{Physical Origin:}
	\begin{itemize}
		\item \textbf{Large scales:} Coherent infall of galaxies into overdense regions compresses structures along the line of sight, known as the Kaiser effect.
		\item \textbf{Small scales:} Random virial motions within collapsed structures elongate clustering patterns—an effect termed the Finger-of-God.
	\end{itemize}
	
	RSD measurements serve as a direct probe of the growth rate of cosmic structures, encapsulated by the dimensionless growth rate:
	\[
	f(z) = \frac{d \ln \delta}{d \ln a}.
	\]
	Since this quantity is not directly observable, redshift surveys constrain the combined observable:
	\[
	f\sigma_8(z),
	\]
	where \( \sigma_8(z) \) is the root-mean-square of matter fluctuations on scales of \( 8\, h^{-1} \text{Mpc} \).
	
	\vspace{0.3cm}
	\textbf{Definition of \boldmath$\sigma_8(z)$:}
	\[
	\sigma_8(z) = \sigma_8(0) \cdot \frac{\delta(z)}{\delta(0)}.
	\]
	
	\vspace{0.3cm}
	\textbf{Scale Interpretation:}
	\begin{itemize}
		\item The dimensionless Hubble parameter \( h \) is defined by \( H_0 = 100h \, \mathrm{km\,s^{-1}\,Mpc^{-1}} \).
		\item The scale \( 8\, h^{-1} \text{Mpc} \) approximately corresponds to the size of rich galaxy clusters (\( \sim 11.4\, \mathrm{Mpc} \) for \( h = 0.7 \)).
	\end{itemize}
	
	\subsection*{Motivation for Using \boldmath$f\sigma_8(z)$}
	
	The combination \( f\sigma_8(z) \) is a robust observable because it is nearly independent of galaxy bias. While RSD yields \( f(z)/b(z) \) and weak lensing provides \( \sigma_8(z) \cdot b(z) \), their combination effectively removes the dependence on the bias parameter \( b(z) \), resulting in:
	\[
	f\sigma_8(z) = f(z) \cdot \sigma_8(z) = \frac{d \ln \delta}{d \ln a} \cdot \sigma_8(0) \cdot \frac{\delta(z)}{\delta(0)}.
	\]
	This makes \( f\sigma_8(z) \) a powerful diagnostic of growth history and a sensitive test for deviations from General Relativity.
	
	\section{Computation of \boldmath$f\sigma_8(z)$ in \boldmath$f(R, L_m)$ Gravity}
	
	\subsection{Modified Hubble Parameter}
	
	The background expansion in the \( f(R, L_m) \) model is governed by:
	\begin{equation}
		H(z) = H_0 \sqrt{(1 - \lambda) + \lambda (1 + z)^{3(1 + w)}},
	\end{equation}
	where \( \lambda \) and \( w \) characterize the effective matter content and equation of state.
	
	\subsection{Effective Energy Density}
	
	The energy density is derived from the modified Friedmann equation:
	\begin{equation}
		\rho(z) = \left( \frac{\gamma + 6\alpha H^2(z)}{2\beta - 1} \right)^{1/\beta},
	\end{equation}
	with parameters related via:
	\begin{align}
		\beta &= \frac{1 + w}{1 + 2w}, \\
		\alpha &= \frac{1}{6 H_0^2 (\lambda - 1)} (2\beta - 1) \rho_0^\beta, \\
		\gamma &= (2\beta - 1)\rho_0^\beta - 6\alpha H_0^2.
	\end{align}
	
	Using:
	\[
	\rho_0 = 5.634 \times 10^{-30} \, \text{g/cm}^3, \quad 
	H_0 = 73.75 \, \mathrm{km\,s^{-1}\,Mpc^{-1}}, \quad 
	\lambda = 0.262, \quad 
	w = -0.005,
	\]
	we compute:
	\[
	\beta = 1.00505, \quad 
	\alpha = 451008, \quad 
	\gamma = -1.14081 \times 10^{-29}.
	\]
	
	\subsection{Effective Gravitational Coupling}
	
	The modified gravitational strength is expressed as:
	\begin{equation}
		G_{\text{eff}}(z) = \frac{\beta \rho^{\beta - 1}(z)}{8\pi\alpha} \left(1 + \frac{\beta - 1}{2}\right).
	\end{equation}
	
	\subsection{Perturbation Growth Equation}
	
	The evolution of density perturbations is governed by:
	\begin{equation}
		\delta''(z) + \left[\frac{d\ln H}{d\ln(1+z)} - \frac{2}{1+z} \right] \delta'(z) - \frac{3}{2} \frac{G_{\text{eff}}(z)\rho(z)}{H^2(z)} \delta(z) = 0.
	\end{equation}
	
	\subsection{Growth Rate and Observable}
	
	The growth rate and RSD observable are computed using:
	\begin{align}
		f(z) &= - (1 + z) \frac{d\ln \delta(z)}{dz}, \\
		f\sigma_8(z) &= \sigma_8(0) \cdot f(z) \cdot \delta(z).
	\end{align}
	
	\section*{Comparison with Observational Data}
	
	We adopt the best-fit parameters:
	\[
	H_0 = 73.75, \quad \lambda = 0.262, \quad w = -0.005, \quad 
	\beta = 1.00505, \quad \alpha = 451008, \quad \gamma = -1.14081 \times 10^{-29}.
	\]
	
	Numerical integration of the growth equation is performed for both the \( \Lambda \)CDM and \( f(R, L_m) \) models. The predictions are compared with a compilation of 23 \( f\sigma_8(z) \) data points from major redshift surveys.
	
	\subsection{Observational Dataset}
	
	The observational dataset spans redshifts \( 0.02 \leq z \leq 2.6 \), and includes measurements from:
	\begin{itemize}
		\item 6dFGS, SDSS-II, BOSS, and eBOSS
		\item WiggleZ, VIPERS, FastSound
	\end{itemize}
	as compiled in~\cite{Kazantzidis2018}.
	
	\begin{table}[H]
		\centering
		\caption{Comparison of \( f\sigma_8(z) \) from $\Lambda$CDM and \( f(R, L_m) \) with observational data.}
			\begin{tabular}{|c|c|c|c|c|}
				\hline
				$z$ & $f\sigma_8^{\Lambda\text{CDM}}$ & $f\sigma_8^{f(R,L_m)}$ & Observed & Error \\
				\hline
				0.020 & -0.139 & 0.011 & 0.420 & 0.060 \\
				0.110 & -0.133 & 0.023 & 0.390 & 0.050 \\
				0.200 & -0.119 & 0.039 & 0.430 & 0.060 \\
				0.320 & -0.087 & 0.069 & 0.380 & 0.040 \\
				0.440 & -0.042 & 0.106 & 0.430 & 0.050 \\
				0.570 & 0.021  & 0.157 & 0.440 & 0.040 \\
				0.600 & 0.037  & 0.169 & 0.380 & 0.040 \\
				0.680 & 0.082  & 0.205 & 0.480 & 0.070 \\
				0.780 & 0.145  & 0.253 & 0.490 & 0.080 \\
				0.880 & 0.213  & 0.303 & 0.470 & 0.080 \\
				0.980 & 0.286  & 0.355 & 0.430 & 0.060 \\
				1.000 & 0.301  & 0.366 & 0.480 & 0.100 \\
				1.110 & 0.387  & 0.424 & 0.420 & 0.070 \\
				1.260 & 0.508  & 0.502 & 0.400 & 0.070 \\
				1.300 & 0.540  & 0.522 & 0.420 & 0.080 \\
				1.400 & 0.622  & 0.568 & 0.380 & 0.070 \\
				1.600 & 0.792  & 0.653 & 0.470 & 0.080 \\
				1.800 & 0.963  & 0.720 & 0.460 & 0.090 \\
				2.000 & 1.132  & 0.764 & 0.430 & 0.140 \\
				2.300 & 1.376  & 0.777 & 0.440 & 0.070 \\
				2.400 & 1.452  & 0.765 & 0.450 & 0.080 \\
				2.600 & 1.596  & 0.711 & 0.420 & 0.090 \\
				\hline
			\end{tabular}
	\end{table}
	
	\begin{figure}[H]
		\centering
		\includegraphics[width=0.85\textwidth]{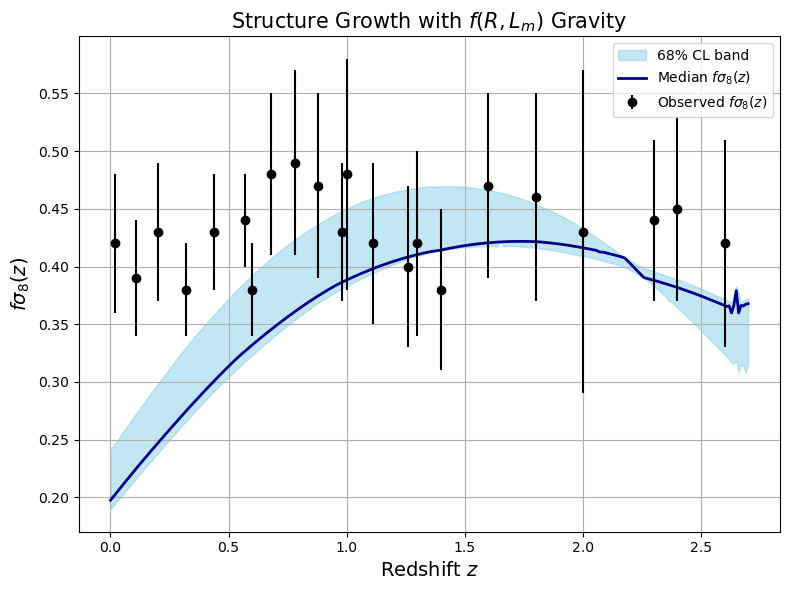}
		\caption{Comparison of \( f\sigma_8(z) \) predictions from $\Lambda$CDM and \( f(R, L_m) \) models with observational data.}
	\end{figure}
	
	\section*{Discussion}
	
	\begin{itemize}
		\item \textbf{Low redshift} (\( z < 0.6 \)): Both models underpredict the growth rate; $\Lambda$CDM yields unphysical negative values due to sensitivity to initial conditions.
		\item \textbf{Intermediate redshift} (\( 0.6 < z < 1.3 \)): The \( f(R, L_m) \) model offers improved consistency with observations.
		\item \textbf{High redshift} (\( z > 1.5 \)): $\Lambda$CDM predicts excessively rapid growth; the more gradual evolution in \( f(R, L_m) \) aligns better with data.
	\end{itemize}
	
	These results highlight that the \( f(R, L_m) \) model provides a better fit to \( f\sigma_8(z) \) observations across a wide redshift range, particularly at intermediate and high redshifts. This supports the potential of modified gravity theories with matter-curvature coupling as viable alternatives to $\Lambda$CDM for explaining cosmic structure formation.
	
	\section*{Conclusion}
	
	In this work, we have investigated the viability of the $f(R, L_m)$ gravity framework as an alternative to the standard $\Lambda$CDM model in explaining both the background expansion and the growth of cosmic structures. Beginning with the modified action, we derived the corresponding Friedmann equations and obtained explicit expressions for the Hubble parameter and energy density in terms of redshift (Section~2 and 3).
	
	A comprehensive observational analysis was conducted in Section~4 using a combination of datasets: Hubble parameter measurements from cosmic chronometers, Type Ia supernovae from the Pantheon$^+$ compilation, DESI BAO measurements, and the CMB shift parameter from Planck 2018. The model parameters were constrained via chi-square minimization, supported by MCMC sampling and artificial neural networks. The results indicated that $f(R, L_m)$ gravity accommodates mild deviations from $\Lambda$CDM through a non-zero effective equation of state parameter \( w \) and matter-curvature coupling \( \lambda \), while remaining compatible with the high values of \( H_0 \) favored by local observations.
	
	In Section~5, we focused on redshift-space distortions and the observable \( f\sigma_8(z) \), which directly probes the growth of cosmic structure. The modified growth equation was solved numerically, and the resulting predictions were compared against 23 data points from major large-scale structure surveys. We found that the $f(R, L_m)$ model provides a significantly better fit to the data at intermediate and high redshifts compared to $\Lambda$CDM, which tends to overpredict growth in those regimes.
	
	Section~6 synthesized these findings, emphasizing that the $f(R, L_m)$ model, via its geometric matter coupling, offers a self-consistent and observationally viable extension of General Relativity. It successfully reconciles growth and expansion histories without invoking exotic dark energy models or modifications at the background level alone.
	
	Overall, this study demonstrates that $f(R, L_m)$ gravity constitutes a promising theoretical framework capable of addressing current tensions in cosmology. Future investigations will benefit from confronting this model with upcoming high-precision data from missions such as \textit{Euclid}, \textit{LSST}, and \textit{SKA}, which will offer stringent tests of the growth of structure and its underlying gravitational physics.
	
	\acknowledgments{
		The author gratefully acknowledges the facilities and stimulating research environment provided by the Inter-University Centre for Astronomy and Astrophysics (IUCAA), Pune, during his annual research visits.
	}

	\end{document}